# Free Fall in Gravitational Theory

W. Engelhardt
Fasaneriestrasse 8, D-80636 München, Germany
wolfgangw.engelhardt@t-online.de

**Abstract**
Einstein's explanation of Mercury's perihelion motion has been verified by astronomical observations. His formula could also be obtained in Schwarzschild metric and was published already in 1898. Motion along a straight geodesic, however, namely free fall into a gravitational centre with vanishing angular momentum, is incorrectly described both by Einstein's and by Schwarzschild's equation of motion. A physical solution for free fall may be obtained by taking into account the dependence of mass on velocity in Newton's gravitational law as adopted in the physics of accelerators.

**Résumé**
L'explication que donne Einstein sur le mouvement de la périhélie de Mercure a été vérifiée grâce à des observations astronomiques. L'on pourrait également obtenir en métrique de Schwarzschild sa formule qui fut déjà publiée en 1898. Cependant, le mouvement tout au long d'une géodésie en ligne droite, à savoir en chute libre vers un centre gravitationnel sans moment cinétique, est décrit de manière incorrecte dans les deux cas: par l'équation de mouvement d'Einstein et par celle de Schwarzschild. Une solution physique pour la chute libre peut s'obtenir en tenant compte du fait que la masse dépend de la vélocité dans la loi sur la gravitation de Newton comme elle est adoptée dans la physique des accélérateurs.

**I Introduction**

In 1915 Einstein published his famous paper on the explanation of Mercury's perihelion motion [1] as a first application of his new gravitational theory. He arrived at the identical formula for the advance of the perihelion which Gerber had derived in 1898 [2] on the basis of a velocity dependent gravitational potential. The velocity dependence chosen by Einstein was somewhat different, but in the approximation considered his result agreed with Gerber's. Since then the accordance between the Gerber formula and the astronomical observations on Mercury's orbit is considered as a corner stone confirming Einstein's geometrized theory of gravitation (GR) that was published in 1916 [3] quoting the foregoing paper of 1915.

In this paper we will show in Sec. 2 that Einstein actually did not use his GR equation of motion, but used Newton's equation with a slight modification of the gravitational potential when he calculated Mercury's anomalous orbit. Thus, his theory suffers from the same deficiency as Newton's, namely free falling mass points may attain superluminal velocities depending on their initial velocity at infinity.

This problem does not arise in the Schwarzschild solution of 1916 [4] which we discuss in Sec. 3. It results in an equation of motion which is different from Einstein's, but it leads also to the Gerber formula. In two letters to Einstein [5] and to Sommerfeld [6] Schwarzschild called it a "miracle" that his abstract idea leads to the same practical result published by Gerber in 1898 and by Einstein in 1915. If one applies, however, Schwarzschild's equation of motion to the free fall of a mass point with vanishing angular momentum, one finds that the kinetic energy of the mass point decreases and the particle comes to a halt at the Schwarzschild radius. This solution avoids superluminality, but it is not physical either.

In Sec. 4 we modify Newton's equation of motion by introducing the velocity dependent mass that proved appropriate in particle accelerators and prevents superluminality in free fall. The particle's kinetic energy increases unlimited when it falls into the assumed singularity at r = 0. The consequences for closed orbits are discussed. One obtains a perihelion advance which is one third of the Gerber-Einstein formula.

**II Einstein's perihelion formula of 1915**
In § 2 of his perihelion paper[a] [1] Einstein derived the equations of motion (7 E) of a mass point in the gravitational field. In lowest order the equations are identical with Newton's, but in higher order Einstein finds from his theory the equations of motion:

$$\frac{d^2 x_\nu}{ds^2} = -\frac{\alpha}{2}\frac{x_\nu}{r^3}\left(1+\frac{\alpha}{r}+2u^2-3\left(\frac{dr}{ds}\right)^2\right) \tag{7b E}$$

Together with the exact conservation of angular momentum

$$r^2 \frac{d\phi}{ds} = B \tag{10 E}$$

and the definition

$$u^2 = \left(\frac{dr}{ds}\right)^2 + r^2\left(\frac{d\phi}{ds}\right)^2 \tag{8a E}$$

one obtains from (7b E)

$$\frac{d^2 x_\nu}{ds^2} = -\frac{\alpha}{2}\frac{x_\nu}{r^3}\left(1+\frac{\alpha}{r}-u^2+\frac{3B^2}{r^2}\right)$$

Einstein claims that this result could be written in the form:

$$\frac{d^2 x_\nu}{ds^2} = \frac{1}{2}\frac{\partial}{\partial x_\nu}\left(\frac{\alpha}{r}+\frac{\alpha B^2}{r^3}\right) = -\frac{x_\nu}{2}\left(\frac{\alpha}{r^3}+\frac{3\alpha B^2}{r^5}\right) \tag{7c E}$$

but this is not the case except under the condition

$$u^2 = \frac{\alpha}{r}$$

It refers to vanishing total energy $A=0$ corresponding to a parabolic orbit. Einstein's equation

$$\left(\frac{dx}{d\phi}\right)^2 = \frac{2A}{B^2}+\frac{\alpha}{B^2}x-x^2+\alpha x^3 \tag{11 E}$$

which he derived from his (7c E) refers, however, to a closed orbit. It is not a consequence of his geometrized gravitational law, but follows from Newton's equation of motion with a modified velocity dependent potential

$$\Phi = -\frac{\alpha}{2r}\left(1+\frac{v_\phi^2}{c^2}\right)$$

---

[a] The title of this paper reads: *Erklärung der Perihelbewegung des Merkur aus der allgemeinen Relativitätstheorie (Explanation of the Perihelion Motion of Mercury from General Relativity Theory)*. As will be shown in Sec. II and in the Appendix, this title is inappropriate, since Einstein's derivation of Gerber's formula is flawed. His "explanation" is not valid.

When he calculated the perihelion advance from (11 E) he miscalculated the last integral, but he reformulated his final result

$$\varepsilon = 24\pi^3 \frac{a^2}{T^2 c^2 (1-e^2)} \qquad (14\ E)$$

such that it was identical with Gerber's formula. It is difficult to retrace why he transformed his direct and simpler result

$$\varepsilon = 3\pi \frac{\alpha}{a(1-e^2)} \qquad (13\ E)$$

by introduction of the orbital period $T$ into the form (14 E) which was not known to him according to his own assertion.

For the case $B = 0$ which corresponds to free fall of a mass point with vanishing angular momentum, Einstein's equation of motion (7c E) is identical with Newton's and suffers from the same deficiency. For an initial radial velocity at infinity very close to the velocity of light one obtains from the Newtonian energy equation

$$\frac{v_r^2}{c^2} = \frac{v_\infty^2}{c^2} + \frac{\alpha}{r} \qquad (1)$$

resulting in superluminal velocities for cosmic radiation particles, for example. This is a result of cancelling the mass in Newton's law so that the velocity dependence is ignored. We come back to this problem in Sec. 4. Before, we discuss the Schwarzschild solution and its consequence for planetary motion.

**III Planetary motion in Schwarzschild metric**

Einstein's equation $R_{ik} = 0$ in empty space is satisfied by Schwarzschild's line element [7]:

$$ds^2 = \frac{1}{K} dr^2 + r^2 d\theta^2 + r^2 \sin^2\theta\, d\phi^2 - K c^2 dt^2, \quad K = 1 - \frac{2GM}{rc^2} = 1 - \frac{\alpha}{r} \qquad (2)$$

when the gravitational field is produced by a point mass M, where G is Newton's gravitational constant and $\alpha$ is twice the so called Schwarzschild radius. For motion in the plane $\theta = \pi/2$ one has to solve the following equations:

$$\frac{d^2\phi}{ds^2} + \frac{2}{r}\frac{dr}{ds}\frac{d\phi}{ds} = 0 \qquad (3)$$

with the integral

$$r^2 \frac{d\phi}{ds} = \frac{B}{i} \qquad (4)$$

reflecting the conservation of angular momentum, and

$$\frac{d^2 t}{ds^2} + \frac{\alpha}{K}\frac{1}{r^2}\frac{dr}{ds}\frac{dt}{ds} = 0 \qquad (5)$$

with the integral

$$\frac{dt}{ds} = \frac{C}{i c K} \qquad (6)$$

Conservation of energy follows from (2) in the form:

$$\frac{1}{K}\left(\frac{dr}{ds}\right)^2 + r^2 \left(\frac{d\phi}{ds}\right)^2 - K c^2 \left(\frac{dt}{ds}\right)^2 = 1 \qquad (7)$$

Substituting the integrals (4) and (6) one obtains with

$$\frac{dr}{ds} = \frac{dr}{d\phi}\frac{d\phi}{ds} \text{ and } r = 1/x$$

Einstein's equation (11 E) if one substitutes for the constant $C^2 = 1 + 2A$. Integrating this equation correctly one obtains in lowest order of $\alpha$ Gerber's formula (14 E) as given by Einstein.

It was this fortunate agreement that caused Schwarzschild to speak of a "miracle". If one calculates, however, the motion along a straight geodesic in the Schwarzschild metric, namely free fall into a gravitational center with vanishing angular momentum, one obtains a result very different from Einstein's. Equation (7) becomes with $dr/dt = v_r$:

$$\frac{v_r^2}{c^2} = K^2\left(1 - \frac{K}{C^2}\right) \tag{8}$$

The integration constant $C$ may be expressed by the initial velocity $v_\infty^2$ at infinity where $K = 1$ so that the radial velocity becomes:

$$\frac{v_r^2}{c^2} = \left(1 - \frac{\alpha}{r}\right)^2 \left[\frac{\alpha}{r}\left(1 - \frac{v_\infty^2}{c^2}\right) + \frac{v_\infty^2}{c^2}\right] \tag{9}$$

As the mass point approaches the Schwarzschild radius, the velocity tends to zero and the falling mass comes to a halt with vanishing kinetic energy. This is not a physical solution, since experience tells us that falling objects gain kinetic energy continuously.

We must conclude that neither Einstein's nor Schwarzschild's version of a GR-solution for the motion on a straight geodesic is in agreement with known facts. It is apparently necessary to take into account the velocity dependent mass as it is applied in accelerator physics. This will be considered in Sec. IV.

**IV Free fall with velocity dependent mass**
Newton's law with velocity dependent mass reads:

$$\frac{d}{dt}(m\vec{v}) = -m\nabla\Phi, \quad m = \frac{m_0}{\sqrt{1 - v^2/c^2}} \tag{10}$$

Differentiation and scalar multiplication with $\vec{v}$ yields:

$$\frac{1}{1 - v^2/c^2}\frac{dv^2}{dt} = \vec{v}\cdot\nabla\left(\frac{\alpha}{r}\right) = \frac{dr}{dt}\frac{\partial}{\partial r}\left(\frac{\alpha}{r}\right) = \frac{d}{dt}\left(\frac{\alpha}{r}\right) \tag{11}$$

Integrating with respect to time one obtains in Einstein's nomenclature:

$$1 - \frac{v^2}{c^2} = \exp\left(-2A - \frac{\alpha}{r}\right) \tag{12}$$

or:

$$1 - \frac{v^2}{c^2} = \left(1 - \frac{v_\infty^2}{c^2}\right)\exp\left(-\frac{\alpha}{r}\right) \tag{13}$$

where the integration constant $A$ has been expressed by the initial velocity at infinity. Equation (13) ensures that the radial velocity in free fall can never exceed the velocity of light. On the other hand, the particle's energy increases indefinitely when the centre of gravity $r = 0$ is approached:

$$E^2 = E_\infty^2 \exp(\alpha/r)$$

Solution (12) disagrees both with the Einstein-Newton solution (1) and with Schwarzschild's result (9), but it is in accordance with the velocity dependent mass increase as established in accelerator physics.

At last we calculate the influence of the velocity dependent mass on the perihelion motion for closed orbits with $A < 0$. From the conservation of angular momentum follows

$$\frac{v_\phi^2}{c^2} = \frac{B^2}{r^2}\left(1 - \frac{v^2}{c^2}\right) = \frac{B^2}{r^2}\exp\left(-2A - \frac{\alpha}{r}\right) \tag{14}$$

and from (12) we have

$$\left(\frac{1}{r^2}\left(\frac{dr}{d\phi}\right)^2 + 1\right)\frac{v_\phi^2}{c^2} = 1 - \exp\left(-2A - \frac{\alpha}{r}\right) \tag{15}$$

Eliminating the tangential velocity we obtain for the orbit with $r = 1/x$:

$$\left(\left(\frac{dx}{d\phi}\right)^2 + x^2\right)B^2 = \exp(2A + \alpha x) - 1 \tag{16}$$

Substituting the constants by the parameters of a Kepler ellipse where $p$ is the semi-latus rectum and $e$ the eccentricity one obtains:

$$\left(\frac{dx}{d\phi}\right)^2 = \frac{2}{\alpha p}\left(\exp\frac{\alpha}{2}\left(\frac{e^2-1}{p} + 2x\right) - 1\right) - x^2 \tag{17}$$

Expanding this for $\alpha = 1$ yields

$$\frac{dx}{d\phi} = \sqrt{-x^2 - \frac{1-e^2}{p^2} + \frac{2x}{p} + \frac{\alpha(1-e^2-2px)^2}{4p^3}} \tag{18}$$

Integration over x between the zeros of $dx/d\phi$ yields the perihelion advance for a full period:

$$\varepsilon = \frac{\alpha \pi}{p} = \frac{\alpha \pi}{a(1-e^2)} \tag{19}$$

This is one third of the Gerber-Einstein formula (13 E). The same result was already obtained by Gerold von Gleich in 1925 [8] [b].

One may wonder why Einstein ignored the variable mass in equation (10) that was proposed by Planck in 1906 [9]. It is often referred to as "relativistic" mass which he certainly would have been aware of in 1915.

**V Conclusions**

Our analysis revealed that the problem of free fall into a gravitational centre is not adequately described in the framework of General Relativity, be it Einstein's or Schwarzschild's version of it. We also found that Einstein actually did not use the equation of motion he had obtained from his geometrized gravitational theory when he derived the formula of Mercury's perihelion advance that was published by Gerber seventeen years before.

---

[b] A profound and learned review of Einstein's relativity theories may be found in:
Gerold von Gleich, *Einsteins Relativitätstheorien und physikalische Wirklichkeit,* Johann Ambrosius Barth, Leipzig 1930. A reprint is in preparation.

In view of these findings it is doubtful whether a geometrized gravitational theory – which is reminiscent of Kepler's laws – is capable of describing the dynamic phenomena due to gravitational forces. In practice only the centre of gravity of a planet could move on a geodesic line in space-time that is practically coincident with a Kepler orbit, but all other mass points are barred from this line by internal tensions. Tidal forces can hardly be described by force-free motion in space-time.

In equilibrium it is obviously necessary to balance gravitational forces with elastic forces that keep objects on the earth's surface, for example. GR does not tell us how the effects of space-time flexion connect to the everyday experience of countless forces. It took centuries to develop the extremely useful and successful physical concept of forces. One should not abandon it light-heartedly.

# Appendix

**Attempt to solve Einstein's equation of motion in order to obtain Gerber's formula**

Einstein's use of the variable *"s"* is not unique in [1]. This may lead to confusion which, however, can be avoided by elimination of this variable and attempting a direct derivation of his equation (11) from his equation of motion (7b).

Einstein finds up to second order:

$$\frac{d^2 x_\nu}{ds^2} = -\frac{\alpha}{2} \frac{x_\nu}{r^3} \left(1 + \frac{\alpha}{r} + 2u^2 - 3\left(\frac{dr}{ds}\right)^2\right) \tag{7b}$$

which yields with the definitions

$$u^2 = \left(\frac{dr}{ds}\right)^2 + r^2\left(\frac{d\phi}{ds}\right)^2 \tag{8a}$$

$$r^2 \frac{d\phi}{ds} = B \tag{10}$$

the equivalent equation

$$\frac{d^2 x_\nu}{ds^2} = -\frac{\alpha}{2} \frac{x_\nu}{r^3} \left(1 + \frac{\alpha}{r} - u^2 + \frac{3B^2}{r^2}\right) \tag{7b}$$

Scalar multiplication with the components $dx_\nu/ds$ and summation yields:

$$\frac{1}{2}\frac{du^2}{ds} = \frac{\alpha}{2} \frac{d}{ds}\left(\frac{1}{r}\right)\left(1 + \frac{\alpha}{r} - u^2 + \frac{3B^2}{r^2}\right)$$

With $x = 1/r$ and elimination of *s* one obtains a differential equation of first order:

$$\frac{du^2}{dx} = \alpha\left(1 + \alpha x - u^2 + 3B^2 x^2\right) \tag{*}$$

From (8a) and (10) we have $u^2/B^2 = (dx/d\phi)^2 + x^2$ so that Einstein's "solution" (11):

$$B^2\left[\left(\frac{dx}{d\phi}\right)^2 + x^2\right] = 2A + \alpha x + \alpha B^2 x^3 \qquad (11)$$

becomes

$$u^2 = 2A + \alpha x + \alpha B^2 x^3 \qquad (**)$$

This expression does not solve the differential equation (*).

While Gerber's formula may be calculated from (11), this equation itself is not derivable from Einstein's equation of motion contrary to the claim in the title of Einstein's famous paper [1].

**References**

[1] A. Einstein, Königlich-Preussische Akademie der Wissenschaften, Sitzungsberichte 1915 (part 2), 831.
[2] P. Gerber, Z. f. Math. u. Phys. **43** (1898) 93.
[3] A. Einstein, Annalen der Physik (ser. 4), **49** (1916) 769.
[4] K. Schwarzschild, Königlich-Preussische Akademie der Wissenschaften, Sitzung vom 3. Februar 1916; Page 189.
[5] K. S., Letter to Einstein, THE COLLECTED PAPERS OF ALBERT EINSTEIN VOLUME **8**, Part A: Correspondence 1914-1917, Page 224, Document 169.
[6] K. S., Letter to Sommerfeld, *Selbstzeugnisse großer Wissenschaftler*, Kultur & Technik, Heft **4**,1987, Verlag Deutsches Museum, München.
[7] H. Bucerius, M. Schneider, *Himmelsmechanik II,* Bibliographisches Institut, Mannheim 1967.
[8] G. von Gleich, Annalen der Physik, Bd. **383** (1925) 498.
[9] M. Planck*,* Verhandlungen Deutsche Physikalische Gesellschaft, **8** (1906) 136.